\documentclass[prl, twocolumn,superscriptaddress,amsmath,amssymb,showpacs,floatfix,preprintnumbers]{revtex4-2}
\usepackage{amsmath}
\usepackage{graphicx}
\usepackage{dcolumn}
\usepackage{color}
\usepackage{physics}
\DeclareGraphicsExtensions{.png .jpg .pdf}

\begin{document}

\title{Spin Torque Generated by Valley Hall Effect in WSe$_2$}

\author{D. J. P. de Sousa}\email{sousa020@umn.edu}
\author{M. J. Sammon}
\affiliation{Department of Electrical and Computer Engineering, University of Minnesota, Minneapolis, Minnesota 55455, USA}
\author{Raseong Kim}
\author{Hai Li}
\author{Ian A. Young}
\affiliation{Components Research, Intel Corporation, Hillsboro, OR 97124 USA}
\author{Tony Low}\email{tlow@umn.edu}
\affiliation{Department of Electrical and Computer Engineering, University of Minnesota, Minneapolis, Minnesota 55455, USA}

\date{ \today }

\begin{abstract}
 Monolayer transition metal dichalcogenides are promising materials for spintronics due to their robust spin-valley locked valence states, enabling efficient charge-to-spin conversion via valley Hall effect with non-equilibrium spins possessing long spin diffusion lengths of hundreds of nanometers. In this work, we show that the injection of a pure valley current, induced by valley Hall effect in a WSe$_2$ monolayer, imparts a spin torque on the magnetization of an overlaid Fe or CoFe in a tunneling structure. The torque efficiency is found to be comparable to that in conventional perpendicular magnetic tunnel junctions and can be further optimized with valley Hall angle in WSe$_2$. The valley nature of the spin torque gives rise to out-of-plane damping-like torques in a current-in-plane configuration, vanishing charge transport perpendicular-to-the-plane as well as torque efficiency tunable through gating. 

\end{abstract}


\maketitle

\emph{Introduction.} 
The ability to electrically manipulate the magnetization of a ferromagnetic thin film with perpendicular magnetic anisotropy in an efficient manner is envisioned to enable unprecedented technological advances through the implementation of in-memory computing technologies~\cite{ref1, ref2, ref3, delin, refinmemory1, refinmemory2}. Deterministic field-free switching of a perpendicular magnetization requires incident spin currents with a non-vanishing spin component aligned with the perpendicular anisotropy axis. State-of-the-art approaches to the generation of such out-of-plane spins include the charge-to-spin conversion through unconventional spin Hall effects in low-symmetry non-magnets~\cite{ref4, ref5, refWTe2} and the injection of spins by additional ferromagnet layers in fully perpendicular magnetic tunnel junctions~\cite{ref1, SciRep, ref2020, refsun, refsun2}. Recently, a new subfield of spintronics has emerged~\cite{new1, new2, new3, new4} which exploits the valley degree of freedom of electrons in two-dimensional materials. Here, we examine the spin torque generated through a valley polarization on an overlaid ferromagnet, and its efficiency in switching a perpendicular magnetization. Addressing this question is key in evaluating the potential of valley physics for spintronics. 

In monolayer transition-metal dichalcogenides (TMDs), the spin-orbit interaction induces a spin-valley locking, with $\rm{K}$ and $\rm{K}'$ valleys supporting opposite spin states as required by time-reversal symmetry. Spin-resolved photoemission spectroscopy measurements have revealed valley-dependent out-of-plane spin polarized valence states in TMDs~\cite{ref6, ref7, ref8, ref9, ref10}. Subsequently, transport experiments unambiguously demonstrated that a flow of out-of-plane spins in $\rm{WSe}_2$ monolayers can be electrically generated through the valley Hall effect (VHE)~\cite{ref11}, a topological Hall response driven by the finite and opposite Berry curvatures in the two valleys~\cite{ref11, ref12, ref13, ref14, EBarre}. In these reports, an out-of-plane spin/valley polarization of 70\% was observed at the edges of p-doped monolayer $\rm{WSe}_2$ due to VHE~\cite{EBarre}. A similar electrically generated interfacial spin polarization of 38\% was also observed in $\rm{WSe}_2$/graphene heterostructure~\cite{ref11}. Spin-valley locked states in TMDs are also long-lived~\cite{sv1, sv2} with large out-of-plane spin diffusion lengths of hundreds of nanometers~\cite{ref13}. These features indicate the potential of spin-valley locking physics to enable unprecedented applications in spintronics.

In this letter, we address the feasibility of utilizing the spin-valley locking of TMDs to induce reversal of an adjacent perpendicular magnetization. By performing transport calculations on a TMD/insulator/ferromagnetic tunnel junction, we show the existence of a sizable spin torque (ST) acting on the magnetization of a $\rm{Fe}$ or $\rm{CoFe}$ slab originating solely from a non-equilibrium valley polarization in a $\rm{WSe}_2$ monolayer. We found that the torque efficiency due to spin-valley locking depends sensitively on the $\rm{WSe}_2$ doping levels and on the lattice misalignment with the ferromagnet and is comparable to that in perpendicular magnetic tunnel junction. The fact that the ST arises from the flow of out-of-plane spins constitutes a novel mechanism suitable for switching a perpendicular magnetization. Our findings point toward the utilization of spin-valley locking physics of TMDs in novel efficient quantum spintronic devices.

\begin{figure}[t]
\includegraphics[width =\linewidth]{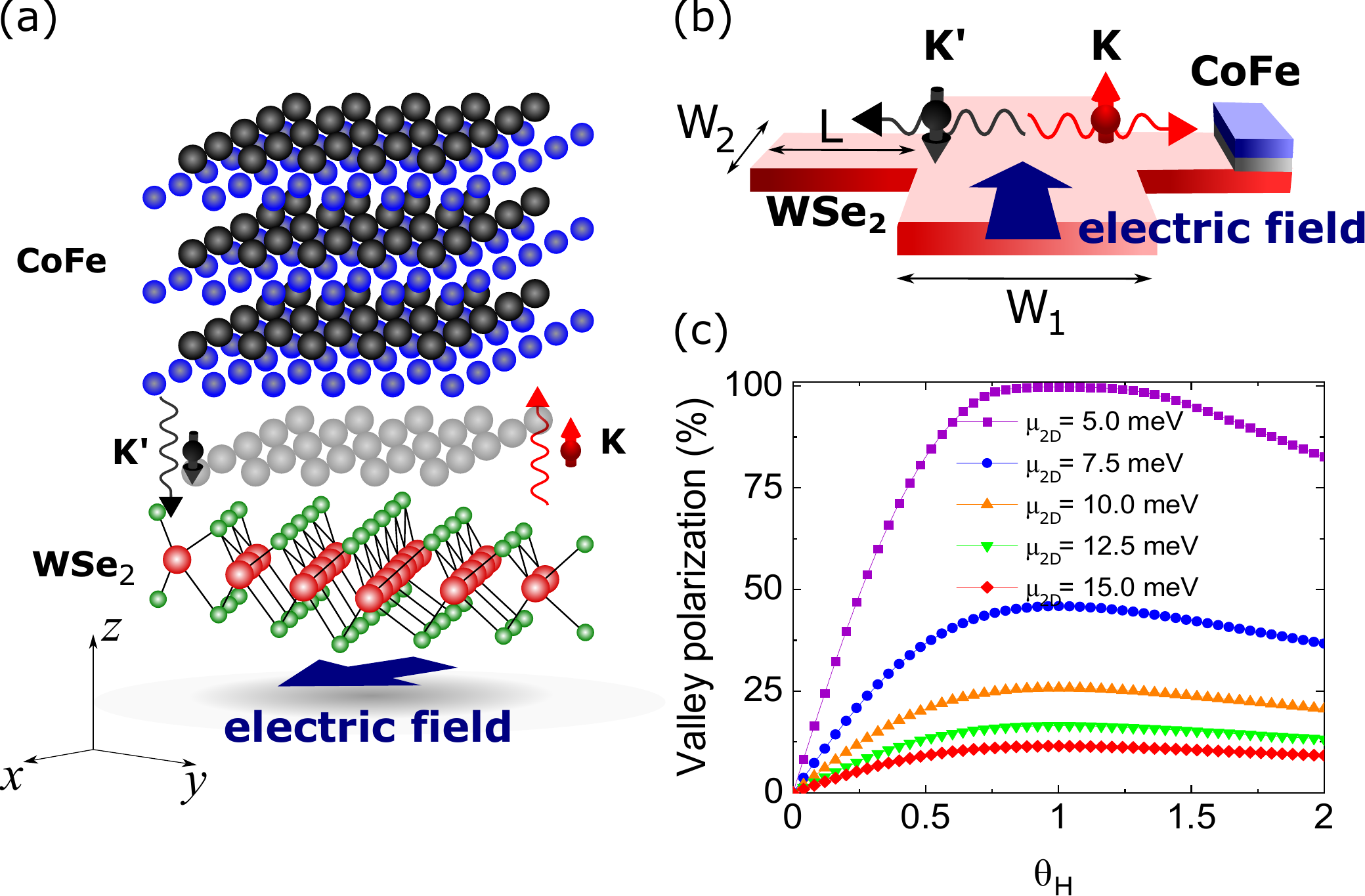}
\caption{(a) Representative $\rm{WSe}_2$/insulator/$\rm{CoFe}$ tunnel junction with vertical tunneling valley/spin currents induced by an in-plane electric field in the WSe$_2$ monolayer. (b) The valley Hall effect causes a non-equilibrium valley imbalance in the transverse side arms of a WSe$_2$ crossbar in response to a longitudinal in-plane electric field. The induced valley voltage is optimized when the dimensions $W_1$, $W_2$ and $L$ are smaller than the spin/valley diffusion length in WSe$_2$. (c) Field-induced non-equilibrium valley polarization in WSe$_2$ as a function of the valley Hall angle at several equilibrium WSe$_2$ doping levels $\mu_{2D}$.}
\label{fig1}
\end{figure}

\emph{Theoretical method.} Figure~\ref{fig1}(a) displays a $\rm{WSe}_2$/insulator/$\rm{CoFe}$ stack as a prototypical example of the tunneling junction studied in this work. In this system, a non-equilibrium valley population induced in the WSe$_2$ portion underneath the CoFe slab gives rise to vertical tunneling valley currents. This situation can be realized by positioning the CoFe slab on one of the side arms of the WSe$_2$ crossbar and taking advantage of the VHE to produce a local non-equilibrium valley density [See Fig.~\ref{fig1}(b)]. The associated non-equilibrium valley chemical potential difference, $\delta \mu = \mu_{\rm{K}} - \mu_{\rm{K'}}$, is well described by the drift-diffusion approach~\cite{ref17}, which renders
\begin{eqnarray}
& \displaystyle \delta \mu = \frac{eI}{\sigma_{xx}} \frac{\theta_H}{1 + \theta_H^2},
\label{eq1}
\end{eqnarray}
where $I$ is the electric field-accompanying in-plane charge current and $\theta_H$ the valley Hall angle of $\rm{WSe}_2$, defined as $\theta_H = \sigma_H/\sigma_{xx}$ where $\sigma_H$ and $\sigma_{xx}$ are the valley Hall and longitudinal charge conductivities, respectively. Our drift-diffusion analysis indicates that Eq.~(\ref{eq1}) is the optimized valley potential induced by VHE in a crossbar geometry. The optimization takes place in the limit where the crossbar arm dimensions, $W_1$, $W_2$ and $L$, become smaller than the spin diffusion length in WSe$_2$, where $\delta\mu$ becomes independent of the crossbar dimensions. In the supplementary material, we present numerical results showing the validity of Eq.~(\ref{eq1}) and how the optimization comes about in the large spin/valley diffusion length limit~\cite{ref17}. The electric field is assumed to be applied along the $x$ direction and the longitudinal charge conductivity is $\sigma_{xx} = (4e^2/h)(\eta \beta)^{-1} \ln[1 + \exp(\beta \mu_{\textrm{2D}})]$, where $\mu_{\textrm{2D}}$ characterizes the doping levels of WSe$_2$ and $\beta = 1/k_B T$, with $k_B$ being the Boltzmann constant and $T$ the temperature~\cite{refnote1}. The valley polarization is defined as $(n_{K} - n_{K'})/(n_{K} + n_{K'}) \times 100 \%$ with valley-dependent carrier density $n_{K(K')} = (m^{*}/\beta \pi \hbar^2)\ln[1 + \exp(\beta [\mu_{\textrm{2D}} \pm \delta\mu /2])]$ where $m^{*}$ is the effective mass of the WSe$_2$ valence states. 

\begin{figure}[t]
\includegraphics[width =\linewidth]{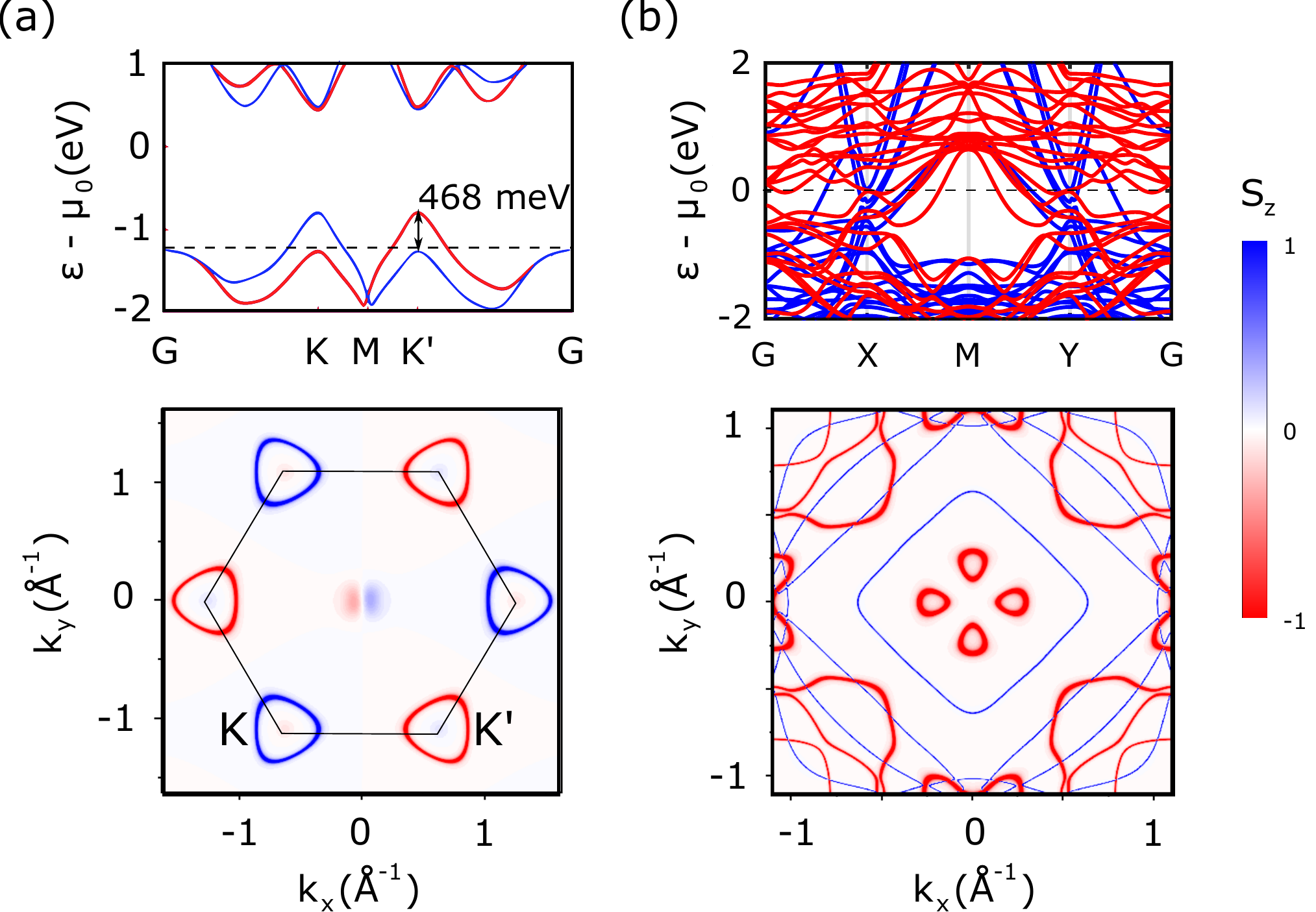}
\caption{The first principles spin-resolved band structure (top panel) and spin density-of-states (bottom panel) of a $\rm{WSe}_2$ monolayer and a 0.7 nm thick $\rm{CoFe}$ slab are shown in panels (a) and (b), respectively. The horizontal dashed line in the band structure plots indicate the energy used in the corresponding spin density-of-states calculations. The valence states of $\rm{WSe}_2$ display a valley-dependent out-of-plane spin polarization ($S_z$) that can be extracted through valley Hall effect~\cite{ref11}.}
\label{fig12}
\end{figure}

Figure~\ref{fig1}(c) reveals that the total electrically generated valley polarization induced at the side arms of the crossbar is a non-monotonic function of $\theta_H$ with a maximum at $\theta_H \approx 1$ for all doping levels, $\mu_{\textrm{2D}}$. Similar non-monotonic behavior was reported in Ref.~\cite{Hai}. Substantial valley polarizations are easily achieved at smaller $\mu_{\textrm{2D}}$ due to the fact that valley depopulation takes place with greater ease at lower $\mu_{\textrm{2D}}$ owing to the larger resistivity. The vertical flow of such non-equilibrium valley-polarized electrons can be addressed through a quantum mechanical tunneling approach, as described in the following.

We employ the Bardeen transfer Hamiltonian formalism~\cite{ref18, ref19} to describe the tunneling process between the $\rm{WSe}_2$ monolayer and ferromagnetic thin film. In Bardeen’s approach, quantum tunneling is treated perturbatively with transition rates being fully described by the electronic ground state of isolated contacts~\cite{ref20, reffreenstra}. The electronic states were obtained with the  pseudopotential/plane-wave method employed in {\fontsize{8}{18}\selectfont QUANTUM ESPRESSO}~\cite{ref15} and subsequently converted to maximally-localized Wannier function basis with the {\fontsize{8}{18}\selectfont WANNIER90} package~\cite{ref16}. Figures~\ref{fig12}(a) and (b) show the spin-resolved band structure and spin density-of-states of a $\rm{WSe}_2$ monolayer and a 0.7 nm thick $\rm{CoFe}$ slab, respectively. The valence states of the $\rm{WSe}_2$ monolayer display giant spin splittings of $\approx 468$ meV, in agreement with previous results~\cite{ref6, ref9}, with nonequivalent valleys $\rm{K}$ and $\rm{K}'$ hosting opposite out-of-plane spin polarizations. This strong spin-valley locking is better appreciated in the bottom panel of Fig.~\ref{fig12}(a), where we show the momentum-resolved spin density-of-states for the energy indicated by the horizontal dashed line in the band structure plot. The Fermi level momentum-resolved spin density-of-states of the $\rm{CoFe}$ slab, shown at bottom panel of Fig.~\ref{fig12}(b), indicates the presence of electron states throughout the whole Brillouin zone including momenta coinciding the $\rm{K}$ and $\rm{K}'$ valleys. These finite vertical tunneling currents require momentum matched states in the two layers. 

\begin{figure}[t]
\includegraphics[width =\linewidth]{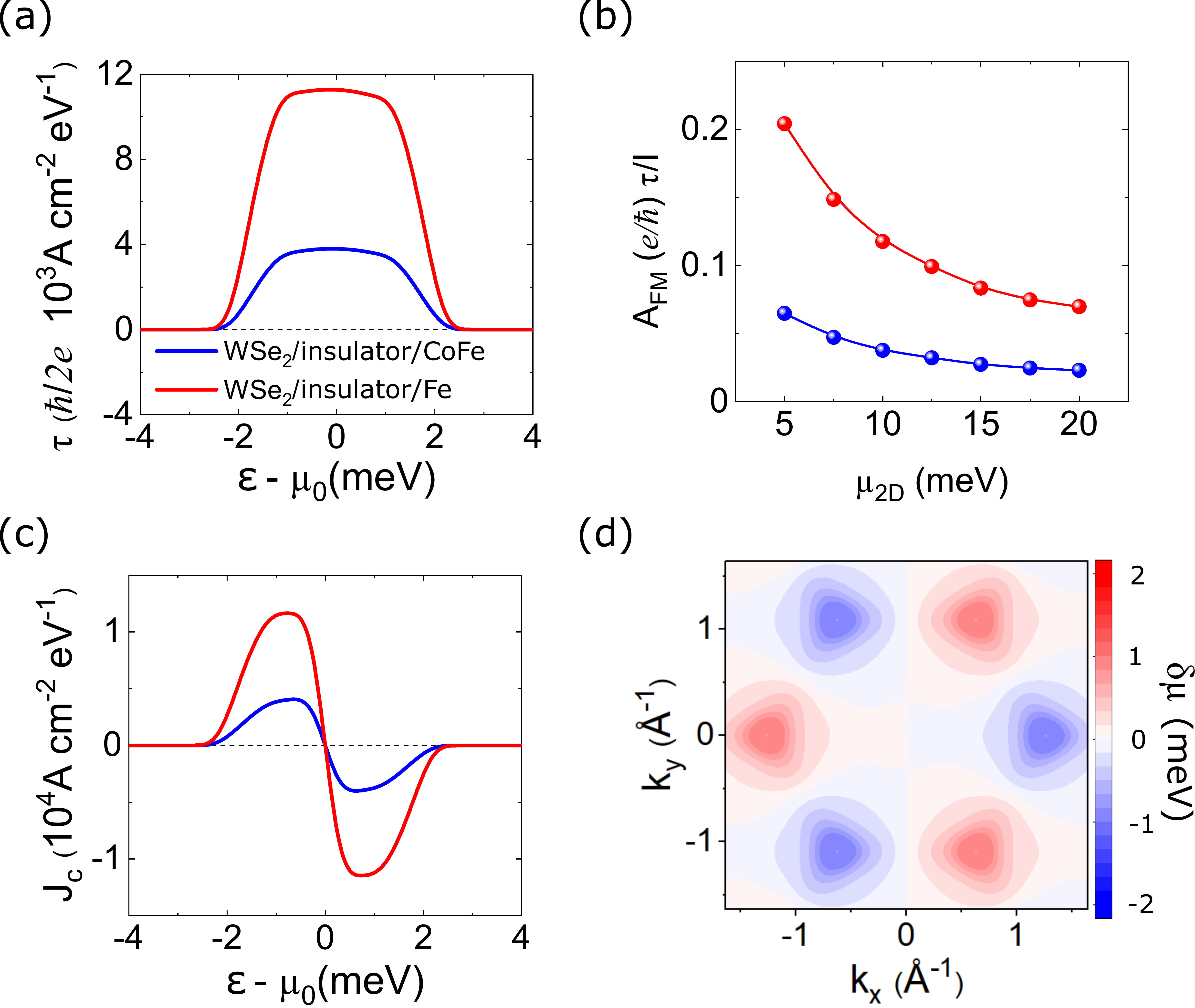}
\caption{(a) Energy-resolved spin torque acting on the magnetization of an $\textrm{Fe}$ (red) and $\textrm{CoFe}$ (blue) slab. The energies are measured in relation to the equilibrium chemical potential of the system $\mu_0$. (b) WSe$_2$ chemical potential dependence of the spin torque. Here, $\mu_{\textrm{2D}}$ characterizes the doping levels of WSe$_2$ only, i.e., the chemical potential of the ferromagnet is kept constant at $\mu_0$ in these calculations. The energy-resolved tunneling charge-current passing through the insulating barrier is shown in panel (c) for both systems. (d) Momentum-resolved non-equilibrium chemical potential of the $\rm{WSe}_2$ monolayer.}
\label{fig2}
\end{figure}

The spin-space components of the tunneling charge-current density reads
\begin{eqnarray}
J^{ss'} = & \nonumber \\ \displaystyle \frac{e}{\hbar}\int d\epsilon \int \frac{d\textbf{k}_{\textrm{T}}}{(2\pi)^2}\int \frac{d\textbf{k}_{\textrm{B}}}{(2\pi)^2}[f(\textbf{k}_{\textrm{T}}) - f(\textbf{k}_{\textrm{B}})] T^{ss'}(\epsilon,\textbf{k}_{\textrm{T}}, \textbf{k}_{\textrm{B}}),
\label{eq2}
\end{eqnarray}
where $f(\textbf{k}_{\textrm{T(B)}})= [1 + \exp(\beta[\epsilon -\mu(\textbf{k}_{\textrm{T(B)}})])]^{-1}$ is the Fermi-Dirac distribution of the top (bottom) electrode and $T^{ss'}(\epsilon,\textbf{k}_{\textrm{T}}, \textbf{k}_{\textrm{B}})$ are the spin space components of the momentum-dependent tunneling rates. The momentum-dependent chemical potentials are $\mu(\textbf{k}_{\textrm{B}}) = \mu_0 + \delta\mu(\textbf{k}_{\textrm{B}})$ and $\mu(\textbf{k}_{\textrm{T}}) = \mu_0$, where $\mu_0$ is the equilibrium chemical potential of the whole structure and $\delta\mu(\textbf{k}_{\textrm{B}})$ equals $+\delta\mu/2$ ($-\delta\mu/2$) at the $\rm{K}$ ($\rm{K}'$) valley due to the VHE. The function $\delta\mu(\textbf{k}_{\textrm{B}})$ is related to the electronic properties of $\rm{WSe}_2$ and applied in-plane current through Eq.~(\ref{eq1}).

The tunneling rate components appearing in Eq.~(\ref{eq2}) are
\begin{eqnarray}
T^{ss'}(\epsilon,\textbf{k}_{\textrm{T}}, \textbf{k}_{\textrm{B}}) = & \nonumber \\ \displaystyle 2\pi \sum_{n,m} |M_{nm}^{ss'}(\textbf{k}_{\textrm{T}}, \textbf{k}_{\textrm{B}})|^2 \delta(\epsilon - \epsilon_{ns\textbf{k}_{\textrm{T}}})\delta(\epsilon - \epsilon_{ms'\textbf{k}_{\textrm{B}}}),
\label{eq3}
\end{eqnarray}
with matrix elements expressed in terms of the electronic eigenstates of isolated top and bottom electrodes through
\begin{eqnarray}
M_{nm}^{ss'}(\textbf{k}_{\textrm{T}}, \textbf{k}_{\textrm{B}}) = & \nonumber \\ \displaystyle \frac{\hbar^2}{2m}\int_{z_0} dS \left( \frac{\partial \psi_{ns\textbf{k}_{\textrm{T}}}}{\partial z}\psi_{ms'\textbf{k}_{\textrm{B}}}^{\dagger} - \psi_{ns\textbf{k}_{\textrm{T}}}\frac{\partial \psi_{ms'\textbf{k}_{\textrm{B}}}^{\dagger}}{\partial z}\right),
\label{eq4}
\end{eqnarray}
where the surface integral is performed halfway inside the barrier at $z_0$ and $m$ is the effective mass of electrons in the insulating region~\cite{ref17}. We use the embedded Green’s function formalism for computing the matrix element~\cite{ref19}. In Ref.~\cite{ref17} we show how this approach can be used to model tunneling through an amorphous tunnel barrier with incommensurate contacts. 

The total tunneling charge-current and out-of-plane polarized spin currents are $J_c = \bar{J} + \Delta J \cos(\theta)$ and $Q^z = \bar{Q}^z + \Delta Q^z \cos(\theta)$, respectively, where $\theta$ is the angle between the magnetization of the ferromagnetic slab and the $z$ axis. We have defined $\bar{J} = (J^{\uparrow \uparrow} + J^{\uparrow \downarrow} + J^{\downarrow \uparrow} + J^{\downarrow \downarrow})/2$, $\Delta J = (J^{\uparrow \uparrow} - J^{\uparrow \downarrow} - J^{\downarrow \uparrow} + J^{\downarrow \downarrow})/2$, $\bar{Q}^z = (\hbar/4e)[J^{\uparrow \uparrow} + J^{\uparrow \downarrow} - J^{\downarrow \uparrow} - J^{\downarrow \downarrow}]$ and $\Delta Q^z = (\hbar/4e)[J^{\uparrow \uparrow} - J^{\uparrow \downarrow} + J^{\downarrow \uparrow} - J^{\downarrow \downarrow}]$. The damping-like torque acting on the magnetization of the ferromagnetic slab is $\boldsymbol{\tau} = \tau \hat{\textbf{m}}\times (\hat{\textbf{m}} \times \hat{\textbf{z}})$ where $\hat{\textbf{m}}$ is the unit vector along the magnetization direction and $\tau = (Q^z(0) - Q^z(\pi))/2 \rightarrow \tau = \Delta Q^z$~\cite{ref21}. In the following, we discuss the main features of the ST as well as the tunneling charge and spin current originating from a valley polarization in the $\textrm{WSe}_2$.

\emph{Results and discussions.} We initially take the zigzag direction of the $\rm{WSe}_2$ monolayer to be aligned with the [100] direction of the $\rm{Fe}$ and $\rm{CoFe}$ slabs, an assumption that will be relaxed later. We assume $\theta_H = 0.4$ throughout~\cite{ref13}. Figure~\ref{fig2}(a) displays the energy dependence of the damping-like torque acting on the magnetization of a $\rm{Fe}$ (red curve) or $\rm{CoFe}$ (blue curve) slab. We have taken the in-plane charge current to be $I = 1 \mu$A and the ST was evaluated at $\theta = \pi/2$, since its angular dependence is fully predetermined. As shown, all contributions to the ST take place within a small energy window around the equilibrium chemical potential, where states with energy above and below $\mu_0$ contribute positively. Hence, a finite integrated ST acts on the magnetization of the ferromagnetic slab originating solely from the non-equilibrium valley-polarized electrons in the $\rm{WSe}_2$ monolayer. Figure~\ref{fig2}(a) also indicates a relatively stronger (weaker) ST in the $\rm{WSe}_2/insulator/\rm{Fe}(\rm{CoFe})$ tunnel junction, which we will revisit later.

\begin{figure*}[t]
\includegraphics[scale = 0.35]{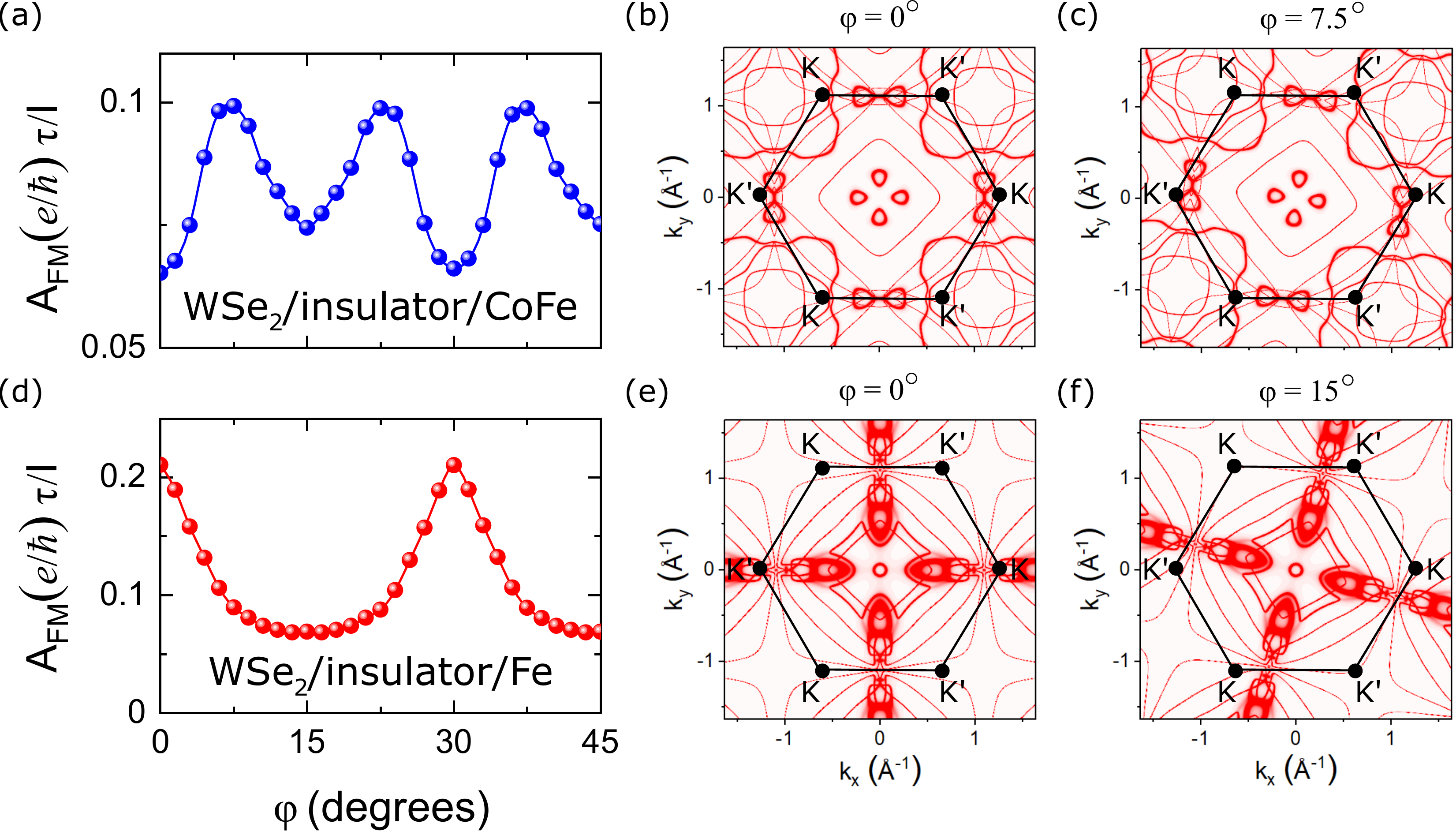}
\caption{Dependence of the spin torque efficiency due to a lattice misalignment. The lattice misalignment is quantified by the angle $\phi$ between the [100] direction of the ferromagnetic layer relative to the zigzag direction of the $\rm{WSe}_2$ monolayer, which is held fixed along the $x$ direction. Panel (a) shows the angular dependence of the torque efficiency in the $\rm{WSe}_2$/insulator/$\rm{CoFe}$ system while panels (b) and (c) show the momentum-resolved density-of-states map of the $\rm{CoFe}$ at $\phi = 0^{\circ}$ and $\phi = 7.5^{\circ}$, respectively. Panels (d), (e) and (f) display similar results but for the $\rm{WSe}_2$/insulator/$\rm{Fe}$ system, where the misalignment angles in the two maps are taken to be $\phi = 0^{\circ}$ and $\phi = 15^{\circ}$. For visualization purposes, we superposed the hexagonal Brillouin zone of $\rm{WSe}_2$, highlighting the valleys.}
\label{fig3}
\end{figure*}

The ST efficiency in a ferromagnetic element of area $\textrm{A}_{\textrm{FM}} \approx 196 \rm{nm} \times 196 \rm{nm}$, defined as $\xi = \textrm{A}_{\textrm{FM}}(e/\hbar)\tau/I$, due to a valley polarization of 70\%  is found to be $\xi_{\rm{CoFe}} \approx 0.07$ and $\xi_{\rm{Fe}} \approx 0.2$ for CoFe and Fe, respectively, for typical $\rm{WSe}_2$ hole concentrations of $p \approx 2.75 \times 10^{11}$ cm$^{-2}$. For the sake of comparison, the spin transfer torque efficiency in perpendicular magnetic tunnel junctions range in the interval $0.01 - 0.1$ with a slight voltage dependence, as estimated from the typical critical switching current density $10^6$~A/cm$^2$ and thermal stability factor of $\approx 60$~\cite{refsun2, refefficiency, footnote2}. Therefore, such valley-induced torque efficiency is large enough to excite magnetization dynamics or reversal of a perpendicular ferromagnetic thin film. We also emphasize that the valley-induced torque efficiency can be further optimized with the valley Hall angle and doping concentrations, as discussed next.  

Our results also indicate that the ST efficiency depends sensitively on the doping levels of $\rm{WSe}_2$, $\mu_{\textrm{2D}}$, as measured from the valence band top. This is shown in Fig.~\ref{fig2}(b) for the $\rm{Fe}$ (red symbols) and $\rm{CoFe}$ (blue symbols) cases. Here, the ST is a monotonically decreasing function of $\mu_{\textrm{2D}}$ with a stronger dependence at lower doping levels for both ferromagnetic slabs. This behavior is a signature of the valley polarization dependence of the ST and can be understood in the low temperature limit as follows: the Fermi-Dirac distribution can approximated as $f(\epsilon - \mu_{\textrm{T(B)}}) \approx \Theta (\mu_{\textrm{T(B)}} - \epsilon)$, where $\Theta$ is the Heaviside function, leading to $\tau \propto \delta \mu$. By explicitly writing $\delta \mu$ in terms of $\mu_{\textrm{2D}}$ through Eq.~(\ref{eq1}) we find $\tau /I \propto 1/\mu_{\textrm{2D}}$, in agreement with the behavior of Fig.~\ref{fig2}(b). This result derives exclusively from the valley physics of $\rm{WSe}_2$ monolayer and suggests that the ST efficiency can be modulated through gating, a feature that has no analogue in conventional spintronics. 

Figure~\ref{fig2}(c) shows that for both ferromagnetic slabs, states with energies above and below $\mu_0$ contribute oppositely to the vertical total charge-current density $J_c$. This implies in the scenario of Fig.~\ref{fig1}(a); The ST acting on the ferromagnet originates from an incoming flow of spin up electrons and an outflow of spin down electrons such that the net tunneling charge current vanishes while a net out-of-plane polarized spin current penetrates the ferromagnet. This is possible by the non-equilibrium valley-dependent chemical potential established in the $\rm{WSe}_2$ monolayer as shown in Fig.~\ref{fig2}(d), where the chemical potential of electrons with momenta at the vicinity of the $\rm{K}$ ($\rm{K}'$) is slightly lower (higher) than that of the ferromagnet, leading to a vertical outflow (inflow) of $\rm{K}$ ($\rm{K}'$) valley-polarized electrons.   

To further explore the nature and behavior of the ST, we study how its magnitude is affected by a lattice misorientation. This is done by rotating the [100] orientation of the ferromagnetic layer by an angle $\phi$ with respect to the zigzag direction of the $\rm{WSe}_2$ monolayer, which is maintained fixed along the $x$ direction. The angular dependence of the torque efficiencies acting on the $\rm{CoFe}$ and $\rm{Fe}$ slabs are shown in Fig.~\ref{fig3}(a) and (d), respectively. The results reveal an oscillatory behavior with a 30$^{\circ}$ period for both cases, where the maximum-to-minimum torque ratio, $\tau_{\textrm{max}}/\tau_{\textrm{min}}$, is approximately $1.5$ and $3$ for the $\rm{CoFe}$ and $\rm{Fe}$ cases, respectively. The stronger angular dependence in the $\rm{Fe}$ case is due to more Fermi surface states at momenta coinciding with the  $\rm{K}$ and  $\rm{K}'$ valleys as compared to $\rm{CoFe}$ system. This is shown in panels (b) and (e) of Fig.~\ref{fig3}, where we show the Fermi level momentum-resolved density-of-states of $\rm{CoFe}$ and $\rm{Fe}$, respectively, at $\phi = 0^{\circ}$. The superposed hexagonal Brillouin zone of the $\rm{WSe}_2$ monolayer highlights the location of $\rm{K}$ and $\rm{K}'$ valleys. As seen in Fig.~\ref{fig3}(e), the two valleys located at $k_y = 0$ \AA\ $^{-1}$ coincide with momentum space density-of-states hot spots of $\rm{Fe}$, giving rise to a larger tunneling spin current. Such Fermi surface matching is less ideal for the $\rm{CoFe}$ system.  Figures~\ref{fig3}(b) and (e) show the hotspot misalignments giving rise to the maximum ST efficiency in CoFe and Fe systems, respectively, while Figs.~\ref{fig3} (c) and (f) display the configurations giving rise to the minimum ST efficiency [See Supplementary material~\cite{ref17} for more discussion].

\emph{Conclusions.} We have shown that a valley Hall effect-induced non-equilibrium valley polarization in a $\rm{WSe}_2$ monolayer results in a spin torque (ST) acting on the magnetization of a ferromagnetic thin film in $\rm{WSe}_2$/insulator/ferromagnetic tunnel junctions. The valley-induced damping-like torque arises from the flow of out-of-plane spin polarized electrons and, therefore, is suitable for exciting magnetization dynamics of thin films with perpendicular magnetic anisotropy, a novel approach that does not require the presence of additional ferromagnetic elements or mirror-broken spin Hall materials. The valley-induced ST was shown to display an efficiency comparable to that in perpendicular magnetic tunnel junctions for typical valley polarizations reported from experiments, with further room for improvement through the valley Hall angle and doping concentrations in the $\rm{WSe}_2$ monolayer, a feature that enables one to modulate the ST efficiency through gating. Finally, we demonstrate how the lattice misalignment and different magnetic layers affect the spin torque efficiency.  

\textit{Acknowledgments}. This material is based upon work supported by Intel Corporation from the University Center (Valleytronics) program. We acknowledge the Minnesota supercomputing institute (MSI) for providing the computational resources. We thank Punyashloka Debashis at Intel Corporation for helpful discussions.


\begin{thebibliography}{apsrev}
\bibitem{refnote1}
In our calculations, we assume $T = 1$ K for the temperature, $\eta$ = 0.1 eV for the spectral broadening and $I = 1 \mu$A for the in-plane electric field-accompanying charge current.
\bibitem{footnote2}
In these estimations, we assume $\alpha = 0.001$ is the intrinsic damping of the ferromagnetic slab.
\end{thebibliography}
\end{document}